\documentclass[floatfix,showpacs,superscriptaddress,prl,twocolumn]{revtex4}
\usepackage{amsbsy,amssymb,amsmath,bm}
\usepackage{epsf}
\usepackage{graphicx,color}
\usepackage{amsfonts}
\usepackage{amssymb}
\usepackage{amsmath}

\input{tcilatex}

\begin{document}

\title{Ballistic transport, chiral anomaly and emergence of the neutral
electron - hole plasma in graphene}
\author{H.C. Kao}
\affiliation{\textit{Physics Department, National Taiwan Normal University, Taipei...,
Taiwan, R. O. C}.}
\author{M. Lewkowicz}
\affiliation{\textit{Applied Physics Department, Ariel University Center of Samaria,
Ariel 40700, Israel}}
\author{B. Rosenstein }
\email{vortexbar@yahoo.com}
\affiliation{\textit{Electrophysics Department, National Chiao Tung University, Hsinchu
30050,} \textit{Taiwan, R. O. C}.}
\date{\today }

\begin{abstract}
The process of coherent creation of particle - hole excitations by an
electric field in graphene is quantitatively described using a dynamic
"first quantized" approach. We calculate the evolution of current density,
number of pairs and energy in ballistic regime using the tight binding
model. The series in electric field strength $E$ up to third order in both
DC and AC are calculated. We show how the physics far from the two Dirac
points enters various physical quantities in linear response and how it is
related to the chiral anomaly. The third harmonic generation and the
imaginary part of conductivity are obtained. It is shown that at certain
time scale $t_{nl}\propto E^{-1/2}$ the physical behaviour dramatically
changes and the perturbation theory breaks down. Beyond the linear response
physics is explored using an exact solution of the first quantized
equations. While for small electric fields the I-V curve is linear
characterized by the universal minimal resistivity $\sigma =\pi /2(e^{2}/h)$%
, at $t>t_{nl}$ the conductivity grows fast. The copious pair creation (with
rate $E^{3/2}$), analogous to Schwinger's electron - positron pair creation
from vacuum in QED, leads to creation of the electron - hole plasma at
ballistic times of order $t_{nl}$. This process is terminated by a
relaxational recombination.
\end{abstract}

\maketitle

22.3.10

\section{I. Introduction}

It has been demonstrated recently that a graphene sheet, especially one
suspended on leads, is one of the purest electronic systems. It became
increasingly evident that electronic mobility in graphene is extremely large
exceeding that in best semiconductor 2D systems \cite{GeimPRL08}. The
scattering of charge carriers in suspended graphene samples of submicron
length is so negligible that the transport is ballistic\cite%
{AndreiNN08,BolotinPRL08}. The ballistic flight time can be estimated as $%
t_{bal}=L/v_{g}\,,$ where $v_{g}$ is the graphene velocity characterizing
the "relativistic" spectrum of graphene near Dirac points and $L$ is the
length of the sample. During the ballistic flight conductivity calculated
neglecting interactions with phonons, ripplons, disorder and between
electrons, etc., are consistent with experiments at least near the Dirac
point at which no carriers are present \cite{Lewkowicz}.

In a simplified model of a single graphene sheet (neglecting scattering
processes and electron interactions) the chemical potential is located right
between the valence and conductance bands and the Fermi "surface" consists
of two Dirac points of the Brillouin zone \cite{Castro}. A lot of effort has
been devoted theoretically to the question of dc and ac transport in pure
graphene due to the surprising fact that the conductivity is finite without
any dissipation process present. A widely accepted "standard" value of the
"minimal dc conductivity" at zero temperature, 
\begin{equation}
\sigma _{1}=\frac{4}{\pi }\frac{e^{2}}{h}\text{,}  \label{sigma1}
\end{equation}%
was calculated using a simplified Dirac model \cite%
{Castro,GusyninPRB06,Ando02,Ziegler06,Katsnelson06,Beenaker} by a variety of
methods. Direct application of the the Kubo formula at zero frequency,
disorder strength, temperature and chemical potential \cite%
{GusyninPRB06,Ziegler06} utilizes certain "regularizations" like the $i\eta $
regulator which is interpreted as an "infinitesimal disorder". The regulator
is removed at the end of the calculation. A more customary application of
the Kubo formula starts with finite frequency. As noted by Ziegler and
others \cite{Ziegler06} the order of limits makes a difference and several
other values different from $\sigma _{1}$ were provided for the \textit{same}
system. The standard value $\sigma _{1}$ is obtained using a rather
unorthodox procedure when the dc limit $\omega \rightarrow 0$ is made 
\textit{before }the zero disorder strength limit $\eta \rightarrow 0$ is
taken. If the order of limits is reversed, one obtains \cite{Ziegler06}:

\begin{equation}
\sigma _{2}=\frac{\pi }{2}\frac{e^{2}}{h}\text{.}  \label{sigma2}
\end{equation}%
When the limit is taken holding $\omega =\eta $ one can even obtain a value
of $\sigma _{3}=\pi \frac{e^{2}}{h}$ \cite{Ziegler06}, thus solving the
"missing $\pi $" problem (the same value was obtained very recently using
yet another regularization in \cite{Beneventano09}). Indeed, at least early
experiments on graphene sheets on $Si$ substrates provided values roughly 3
times larger than $\sigma _{1}$ \cite{Novoselov05}. Recent experiments on
suspended graphene \cite{AndreiNN08} demonstrated that the dc conductivity
is lower, $1.7\sigma _{1}$, as temperature is reduced to $4K$. Hence, while $%
\sigma _{3}$ seems to be inconsistent with experiment, one still faces the
question of what is the proper theoretical value. Since the conductivity of
clean graphene in the infinite sample is a well defined physical quantity,
there cannot be any ambiguity as to its theoretical value. Another
theoretical approach to the problem was the use of Landauer formalism to the
graphene sheet conductance. The conductivity appears as a limiting value of
infinite width $W$ in this approach \cite{Katsnelson06,Beenaker} and one
obtains $\sigma _{1}$. Also resumming the series in disorder by a self
consistent Born approximation gives $\sigma _{1}$ \cite{Ando02}, while other
resummations can lead to a different result \cite{FradkinPRB86}.

In contrast to the controversy with respect to the dc conductivity, both the
experimental and the theoretical situation for the ac conductivity in the
high frequency limit is quite different. The theoretically predicted value
in the Dirac model is $\sigma _{2}$ independent of frequency under condition 
$\omega >>T/\hbar $ \cite{Ando02,Varlamov07}. The Dirac model becomes
inapplicable when $\omega \ $is of order of $\gamma /\hbar \approx 4\cdot
10^{15}Hz$ or larger, where $\gamma $ is the hopping energy of graphene. It
was shown theoretically\ using the tight binding model and experimentally in 
\cite{GeimScience08} that the optical conductivity at frequencies higher or
of order $\gamma /\hbar $ becomes slightly larger than $\sigma _{2}$.
Moreover, in light transmittance measurements at frequencies down to $%
2.5\cdot 10^{15}Hz$\ it was found equal to $\sigma _{2}$ within 4\%. The
tight binding model \textit{does not contain any other time} scales capable
of changing the limiting value of the ac conductivity all the way to $\omega
\rightarrow 0$. Therefore one would expect that the dc conductivity is $%
\sigma _{2}$ rather than $\sigma _{1}$.

As is shown in the present paper, using the dynamical approach (a brief
account of some results was published in \cite{Lewkowicz,Rosenstein10}),
this is indeed the case. The basic physical effect of the electric field is
a coherent creation of electron - hole pairs, mostly, but not exclusively,
near the two Dirac points. To effectively describe this process we develop a
dynamical approach to charge transport in clean graphene using the "first
quantized" approach to pair creation physics similar to that developed in
relativistic physics \cite{Gitman} to describe electron - positron pairs
creation. To better visualize the phenomenon of resistivity without
dissipation directly at zero temperature, doping, frequency and disorder we
describe an experimental situation as closely as possible by calculating
directly the time evolution of the electric current after switching on an
electric field. In this way the use of a rather formal Kubo or Landauer
formalism is avoided and as a result \textit{no regularizations are needed}%
.\ Although we consider an infinite sample, the dynamical approach allows us
to obtain qualitative results for finite samples by introducing time cutoffs
like ballistic flight time. Various other factors determining transport can
be conveniently characterized by time scales like the relaxation time for
scattering of phonons or impurities.

We show in detail that some aspects the linear response physics are not
dominated by the two Dirac points of the Brillouin zone at which the
spectrum is gapless. For example, large contributions (infinite, when the
size of the Brillouin zone $2\pi /a$ ($a=3\mathring{A}$ is the lattice
spacing), is being considered infinite) to the conductivity from the
vicinity of the Dirac points are canceled by contributions from the region
between them. This phenomenon is related to the chiral anomaly in field
theory \cite{Smit}. We analyse the use of massless Dirac (Weyl) model
approximation to the tight binding model and propose an effective chirally
invariant regularization for it.

In addition to the analysis of the linear response, we determin the range of
applicability of the linear response approximation by both calculation of
higher orders in (dc and ac) electric field and solving the model exactly in
the nonlinear electromagnetic response regime. Only the zero chemical
potential case (no net charge) is considered. In this respect the work is
complimentary to an earlier study by Mikhailov \cite{Mikhailov} in which
ballistic nonlinear electromagnetic response to an ac field was calculated
at finite chemical potential using the Boltzmann equation within a
semi-classical approach and major effects we study were omitted. We first
obtain perturbative corrections up to the third order in the electric field $%
E$. At this order qualitatively new phenomena like the third harmonic
generation and the imaginary part of conductivity appear. The calculation of
the corrections allows us to estimate the time scale at which perturbation
theory breaks down. At this scale, $t_{nl}\propto E^{-1/2}$, the physical
behaviour is expected to change qualitatively. Therefore one has to resort
to nonperturbative methods. Certain aspects of nonlinear ac response at zero
chemical potential were also studied recently by Mishchenko \cite%
{MishchenkoPRL09}. In his work disorder (taken into account
phenomenologically) dominates the purely ballistic effects by cutting off
the ballistic times at relaxation time scale before $t_{nl}$ is reached.

Physics of the simplest tight binding model beyond the linear response is
explored via an exact solution of the first quantized equations. It is a
peculiar property of the tight binding model in the dynamical approach, that
the exact solution of the equations for arbitrary momentum $k_{y}$ can be
reduced to that for $k_{y}=0$; the constant electric field being in the $y$
direction. Moreover, the remaining equations, using Floquet theory, can then
be reduced to a recursion relation. We use this property to effectively
calculate the long ballistic flight evolution of various physical quantities
like the current density and the number of created pairs. While for small
ballistic times, $t<t_{nl}$, the conductivity settles at $\sigma _{2}$, at $%
t>t_{nl}$ the current grows linearly. This increase can be explained using
Schwinger's electron - positron pair creation mechanism \cite{Schwinger}.
The pair creation is a two dimensional version of that in QED with the
creation rate proportional to $E^{3/2}$\cite{Cohen}. This, in absence of
relaxation channels (for times below $t_{bal}=L/v_{g}\,$), leads to\ the
creation of a neutral electron - hole plasma at ballistic times of order $%
t_{nl}$. This process cannot continue forever and is terminated by a
relaxational recombination. The applicability of Schwinger's formula for the
electron - hole pairs creation rate was debated over a long time and we set
the limitation on the applicability of this exact formula to graphene.

The rest of the paper is organized as follows. The tight binding model and
the dynamical approach are described in section II. The perturbation theory
in electric field is described in section III. The minimal conductivity is
obtained and the role of states beyond the neighbourhoods of the Dirac
points (and their relation to "axial anomaly" and the Nielsen-Ninomya
theorem) is clarified. The dynamical linear response approach to the ac
field is considered. In Section IV perturbative corrections beyond linear
response (up to third order in $E$) are calculated . The third harmonic and
inductive contributions to conductivity are discussed. The exact solution
for arbitrary field is constructed using the Floquet theory in Section V. It
is used in Section VI to discuss the pair creation rate, conductivity and
speculate about the electron - hole plasma. Finally Section VII contains
discussion and conclusions.

\section{II. The model and the dynamic approach to its physical properties}

\subsection{The tight binding model in an electric field}

Electrons in graphene are described sufficiently accurately for our purposes
by the 2D tight binding model of nearest neighbour interactions \cite{Castro}%
. The Hamiltonian in $\mathbf{k}$ space is

\begin{equation}
\widehat{H}=\sum_{\mathbf{k}}%
\begin{pmatrix}
c_{\mathbf{k}}^{A\dag } & c_{\mathbf{k}}^{B\dag }%
\end{pmatrix}%
H_{\mathbf{k}}%
\begin{pmatrix}
c_{\mathbf{k}}^{A} \\ 
c_{\mathbf{k}}^{B}%
\end{pmatrix}%
\text{,}  \label{H}
\end{equation}%
where 
\begin{equation}
H_{\mathbf{k}}=%
\begin{pmatrix}
0 & h_{\mathbf{k}} \\ 
h_{\mathbf{k}}^{\ast } & 0%
\end{pmatrix}%
;\text{ \ \ }h_{\mathbf{k}}=-\gamma \sum_{\alpha }e^{i\mathbf{k}\cdot 
\mathbf{\delta }_{\alpha }}  \label{H_matrix}
\end{equation}%
with $\gamma \simeq 2.7eV$ being the hopping energy; $\delta _{1}=\frac{a}{3}%
\left( 0,\sqrt{3}\right) $and $\ \delta _{2,3}=\frac{a}{3}\left( \pm \frac{3%
}{2},-\frac{\sqrt{3}}{2}\right) $\ are the locations of nearest neighbours
separated by distance $a\simeq 3\mathring{A}$; and the pseudospin index $A,B$
denotes two triangular sublattices. In most parts of the paper we keep the
function $h_{\mathbf{k}}$ arbitrary. Let us first consider the system in a
constant and homogeneous electric field along the $y$ direction switched on
at $t=0.$ It is described by the minimal substitution, 
\begin{equation}
\mathbf{p}=\hslash \mathbf{k}+\frac{e}{c}\mathbf{A,}  \label{minimal}
\end{equation}%
with vector potential $\mathbf{A}=\left( 0,-cEt\right) $ for $t>0$ ($e>0$).
Later it is generalized to more general fields including the ac field.

\subsection{The dynamical "first quantized" approach.}

Since the crucial physical effect of the field is mainly a coherent creation
of electron - hole pairs (though not always near the Dirac points, see
below), a convenient formalism to describe the pair creation is the "first
quantized" formulation described in detail in \cite{Gitman}. The spectrum
before the electric field is switched on is divided into positive and
negative energy parts describing the valence and conduction band:%
\begin{eqnarray}
H_{\mathbf{k}}\left( E=0\right) u_{\mathbf{k}} &=&-\varepsilon _{\mathbf{k}%
}u_{\mathbf{k}};\text{ \ \ \ \ }u_{\mathbf{k}}=%
\begin{pmatrix}
1 \\ 
-h_{\mathbf{k}}^{\ast }/\varepsilon _{\mathbf{k}}%
\end{pmatrix}%
;  \label{uk} \\
H_{\mathbf{k}}\left( E=0\right) v_{\mathbf{k}} &=&\varepsilon _{\mathbf{k}%
}v_{\mathbf{k}};\text{ \ \ \ \ \ }v_{\mathbf{k}}=%
\begin{pmatrix}
1 \\ 
h_{\mathbf{k}}^{\ast }/\varepsilon _{\mathbf{k}}%
\end{pmatrix}%
;  \label{vk}
\end{eqnarray}%
where $\varepsilon _{\mathbf{k}}=\left\vert h_{\mathbf{k}}\right\vert $. A
second quantized state\ is uniquely characterized by the first quantized
amplitude, 
\begin{equation}
\psi _{\mathbf{k}}\left( t\right) =%
\begin{pmatrix}
\psi _{\mathbf{k}}^{1}\left( t\right) \\ 
\psi _{\mathbf{k}}^{2}\left( t\right)%
\end{pmatrix}%
\text{,}  \label{spinor}
\end{equation}%
which is a "spinor" in the sublattice space. It obeys the matrix
Schroedinger equation in sublattice space 
\begin{equation}
i\hslash \partial _{t}\psi _{\mathbf{k}}=H_{\mathbf{p}}\psi _{\mathbf{k}}%
\text{,}  \label{Schroedinger}
\end{equation}
where $\mathbf{p}$ is defined in Eq.(\ref{minimal}). The initial condition
corresponding to a second quantized state at $T=0$ in which all the negative
energy states are occupied and all the positive energy states are empty is 
\begin{equation}
\psi _{\mathbf{k}}\left( t=0\right) =u_{\mathbf{k}}\text{.}  \label{u}
\end{equation}

A physical quantity is usually conveniently written in terms of $\psi $. We
will be interested mostly in the current density (multiplied by factor $2$
due to spin)

\begin{equation}
J_{y}\left( t\right) =-2e\sum_{\mathbf{k\in }BZ}\psi _{\mathbf{k}}^{\dag
}\left( t\right) \frac{\partial H_{\mathbf{p}}}{\partial p_{y}}\psi _{%
\mathbf{k}}\left( t\right) ,  \label{Jy_def}
\end{equation}%
as well as the energy and the density of the electron - hole pairs. The
summation is over the honeycomb-lattice Brillouin zone, see Fig. 1, in which
two Brillouin zones are outlined. The energy of the electrons is changing
due to the applied electric field (with no dissipation in the ballistic
regime, see however Discussion) and is given by 
\begin{equation}
U\left( t\right) =2\sum_{\mathbf{k}}\psi _{\mathbf{k}}^{\dag }\left(
t\right) H_{\mathbf{p}}\psi _{\mathbf{k}}\left( t\right) \text{.}
\label{U_def}
\end{equation}%
The amplitude of lifting an electron into the conduction band (defined for
the Hamiltonian without the electric field) is $\left\langle \psi \left(
t\right) |v_{\mathbf{k}}\right\rangle $, where the scalar product is defined
by $\left\langle \psi |\phi \right\rangle =\psi _{1}^{\ast }\phi _{1}+\psi
_{2}^{\ast }\phi _{2}$, and consequently the density of pairs reads

\begin{equation}
N_{0}\left( t\right) =2\sum_{\mathbf{k}}\left\vert \left\langle \psi \left(
t\right) |v_{\mathbf{k}}\right\rangle \right\vert ^{2}=2\sum_{\mathbf{k}%
}\left\vert \psi _{1}^{\ast }+\frac{h_{\mathbf{k}}^{\ast }}{\varepsilon _{%
\mathbf{k}}}\psi _{2}^{\ast }\right\vert ^{2}\text{.}  \label{N_unren}
\end{equation}%
Since the Hamiltonian depends on time via the vector potential that shifts
the momentum $\mathbf{p}$, a more useful definition of the density of pairs
taking into account the shift of the momentum will be given in Section III.

\subsection{Units and conventions}

The units of energy, time and distance are defined by the microscopic values 
$\gamma ,$ $t_{\gamma }\equiv \hslash /\gamma $ $\left( \simeq 2.5\cdot
10^{-16}s\right) $ and $a,$ correspondingly. The unit of the electric field
will be $E_{0}\equiv \frac{\gamma }{ea}\simeq 10^{10}V/m$, so that the
dimensionless electric field is $\mathcal{E}=E/E_{0}$ and the unit of
conductivity will be $e^{2}/\hslash $. Effectively one can set $\gamma
=\hslash =a=e=1$. In these units the first quantized equation reads 
\begin{eqnarray}
i\partial _{t}\psi _{1} &=&h_{\mathbf{p}}\psi _{2};  \label{Schroedinger1} \\
i\partial _{t}\psi _{2} &=&h_{\mathbf{p}}^{\ast }\psi _{1}\text{,}  \notag
\end{eqnarray}%
where the tight binding Hamiltonian matrix element $h_{\mathbf{p}}$ takes
the form 
\begin{equation}
h_{\mathbf{p}}=-\exp \left( i\frac{p_{y}}{\sqrt{3}}\right) -b\exp \left( -i%
\frac{p_{y}}{2\sqrt{3}}\right) ,  \label{h_p}
\end{equation}%
where $b=2\cos \left( \frac{p_{x}}{2}\right) ,$ and $p_{y}=k_{y}+A_{y},$ $%
A_{y}=-\mathcal{E}t$ for the dc field. We will use extensively its expansion
in powers of $A_{y}$:

\begin{eqnarray}
\frac{dh_{\mathbf{p}}}{dA_{y}} &\equiv &h_{\mathbf{p}}^{\prime }=-\frac{i}{%
\sqrt{3}}\exp \left( i\frac{p_{y}}{\sqrt{3}}\right) +\frac{ib}{2\sqrt{3}}%
\exp \left( -i\frac{p_{y}}{2\sqrt{3}}\right) ;  \notag \\
\frac{d^{2}h_{\mathbf{p}}}{dA_{y}^{2}} &\equiv &h_{\mathbf{p}}^{\prime
\prime }=\frac{1}{3}\exp \left( i\frac{p_{y}}{\sqrt{3}}\right) +\frac{b}{12}%
\exp \left( -i\frac{p_{y}}{2\sqrt{3}}\right) \text{.}  \label{hprime1}
\end{eqnarray}

For example, the off - diagonal matrix element for the $\mathbf{p}$
component of the current $J_{y}$ is $-h_{\mathbf{p}}^{\prime }$, so that the
current density, using the first quantized approach, is

\begin{equation}
J_{y}=-2\sum_{\mathbf{k}}\psi _{\mathbf{p}}^{\dag }%
\begin{pmatrix}
0 & h_{\mathbf{p}}^{\prime } \\ 
h_{\mathbf{p}}^{\ast \prime } & 0%
\end{pmatrix}%
\psi _{\mathbf{p}}\text{.}  \label{Jy1}
\end{equation}%
The next two sections deal with the perturbative treatment of the electric
field and later (after having shown that it fails at certain time scale) we
will switch to a nonperturbative method.

\section{III. Linear response, the pseudo - diffusive behaviour and the
parity anomaly}

\subsection{Expansion in electric field}

Expanding in the dimensionless electric field strength (assumed for
simplicity homogeneous and constant after switching on the field at $t=0$), $%
\psi _{\mathbf{k}}=e^{i\varepsilon _{\mathbf{k}}t}\left( u_{\mathbf{k}}+%
\mathcal{E}\xi _{\mathbf{k}}+...\right) $, one obtains for the first
correction the following equation:

\begin{equation}
i\partial _{t}\xi _{\mathbf{k}}=%
\begin{pmatrix}
\varepsilon _{\mathbf{k}} & h_{\mathbf{k}} \\ 
h_{\mathbf{k}}^{\ast } & \varepsilon _{\mathbf{k}}%
\end{pmatrix}%
\xi _{\mathbf{k}}-t%
\begin{pmatrix}
0 & h_{\mathbf{k}}^{\prime } \\ 
h_{\mathbf{k}}^{\ast \prime } & 0%
\end{pmatrix}%
u_{\mathbf{k}}\text{.}  \label{eq_corr}
\end{equation}%
Writing the correction as $\xi _{\mathbf{k}}=\alpha _{\mathbf{k}}u_{\mathbf{k%
}}+\beta _{\mathbf{k}}v_{\mathbf{k}}$, it takes the form

\begin{eqnarray}
i\partial _{t}\alpha _{\mathbf{k}} &=&t\varepsilon _{\mathbf{k}}^{\prime };
\label{eq_corr1} \\
i\partial _{t}\beta _{\mathbf{k}} &=&2\varepsilon _{\mathbf{k}}\beta _{%
\mathbf{k}}-t\varepsilon _{\mathbf{k}}I_{\mathbf{k}}^{-}\text{,}  \notag
\end{eqnarray}%
with initial conditions $\alpha _{\mathbf{k}}\left( t=0\right) =1$, $\beta _{%
\mathbf{k}}\left( t=0\right) =0$. We use the abbreviations 
\begin{equation}
\varepsilon _{\mathbf{k}}^{\prime }\equiv \frac{\partial \varepsilon _{%
\mathbf{k}}}{\partial k_{y}}  \label{primes}
\end{equation}%
and 
\begin{equation}
I_{\mathbf{k}}^{-}\equiv \frac{h_{\mathbf{k}}^{\ast }h_{\mathbf{k}}^{\prime
}-h_{\mathbf{k}}h_{\mathbf{k}}^{\ast \prime }}{2\varepsilon _{\mathbf{k}}^{2}%
}\text{.}  \label{Im}
\end{equation}%
The coefficient $\beta _{\mathbf{k}}$ denotes the amplitude of accumulation
of electrons in the conduction band, whereas $\alpha _{\mathbf{k}}$ denotes
the amplitude of remaining in the valence band. Solving the equations one
obtains

\begin{equation}
\alpha _{\mathbf{k}}=-\frac{i}{2}t^{2}\varepsilon _{\mathbf{k}}^{\prime };%
\text{ \ \ }\beta _{\mathbf{k}}=\frac{iI_{\mathbf{k}}^{-}}{4\varepsilon _{%
\mathbf{k}}}\left( 1-e^{-2i\varepsilon _{\mathbf{k}}t}-2i\varepsilon _{%
\mathbf{k}}t\right) \text{.}  \label{correction}
\end{equation}

The expansion for the current, Eq.(\ref{Jy_def}), in the electric field up
to the first order contains the following momentum $\mathbf{k}$
contributions, $J_{y}=\sum_{\mathbf{k}}j_{\mathbf{k}}$:%
\begin{equation}
j_{\mathbf{k}}=j_{\mathbf{k}}^{0}+j_{\mathbf{k}}^{p}+j_{\mathbf{k}}^{d}\text{%
.}  \label{j12}
\end{equation}
The zero order contribution is

\begin{equation}
j_{\mathbf{k}}^{0}=-u_{\mathbf{k}}^{\dag }%
\begin{pmatrix}
0 & h_{\mathbf{k}}^{\prime } \\ 
h_{\mathbf{k}}^{\ast \prime } & 0%
\end{pmatrix}%
u_{\mathbf{k}}=2\varepsilon _{\mathbf{k}}I_{\mathbf{k}}^{+}.  \label{j0}
\end{equation}%
where%
\begin{equation}
I_{\mathbf{k}}^{+}=\frac{h_{\mathbf{k}}h_{\mathbf{k}}^{\ast ^{\prime }}+h_{%
\mathbf{k}}^{\ast }h_{\mathbf{k}}^{\prime }}{2\varepsilon _{\mathbf{k}}^{2}}%
\text{.}  \label{Ip}
\end{equation}%
The correction gives the 'paramagnetic' and 'diamagnetic' contributions to
the current densities:

\begin{eqnarray}
j_{\mathbf{k}}^{p} &=&-2\mathcal{E}u_{\mathbf{k}}^{\dag }%
\begin{pmatrix}
0 & h_{\mathbf{k}}^{\prime } \\ 
h_{\mathbf{k}}^{\ast \prime } & 0%
\end{pmatrix}%
\xi _{\mathbf{k}}+cc  \label{jp} \\
&=&2\mathcal{E}\left( \alpha _{\mathbf{k}}-\alpha _{\mathbf{k}}^{\ast
}\right) \varepsilon _{\mathbf{k}}I_{\mathbf{k}}^{-}=\mathcal{E}\left( I_{%
\mathbf{k}}^{-}\right) ^{2}\left[ 2\varepsilon _{\mathbf{k}}t-\sin \left(
2\varepsilon _{\mathbf{k}}t\right) \right] ;  \notag \\
j_{\mathbf{k}}^{d} &=&\mathcal{E}tu_{\mathbf{k}}^{\dag }%
\begin{pmatrix}
0 & h_{\mathbf{k}}^{\prime \prime } \\ 
h_{\mathbf{k}}^{\ast \prime \prime } & 0%
\end{pmatrix}%
u_{\mathbf{k}}=-2\mathcal{E}t\varepsilon _{\mathbf{k}}I_{\mathbf{k}}^{+}%
\text{.}  \label{jd}
\end{eqnarray}%
Both corrections for a specific momentum $\mathbf{k}$ diverge at large
ballistic times as $t$, however one still has to integrate over the valence
band momenta.

\subsection{Integration over momenta and physical interpretation of \ the
"quasi -Ohmic" behaviour.}

To first order in electric field the current density is%
\begin{equation}
J_{y}=J_{0}+\sigma \mathcal{E},  \label{j}
\end{equation}%
where the zero order contribution can be written as a derivative with
respect to $k_{y}$ of a periodic function

\begin{equation}
J_{0}=-2\sum_{\mathbf{k}}j_{\mathbf{k}}^{0}=4\sum_{\mathbf{k}}\varepsilon _{%
\mathbf{k}}^{\prime }.  \label{J0}
\end{equation}%
It vanishes upon integration, in accordance with the Bloch theorem, since
the Brillouin zone can be chosen in such a way that it exhibits periodicity
in the field ($y$) direction. For example, we can integrate over the
rectangular area shown in Fig.1 containing two Brillouin zones: 
\begin{equation}
\sum_{BZ}=\frac{1}{2}\frac{1}{\left( 2\pi \right) ^{2}}\int_{-2\pi }^{2\pi
}dk_{x}\int_{-2\pi /3^{1/2}}^{2\pi /3^{1/2}}dk_{y}\text{.}  \label{integral}
\end{equation}%
It should be noted that $\int_{-2\pi /3^{1/2}}^{2\pi /3^{1/2}}\varepsilon _{%
\mathbf{k}}^{\prime }dk_{y}=\varepsilon _{k_{x},2\pi /3^{1/2}}-\varepsilon
_{k_{x},-2\pi /3^{1/2}}=0\ $at every $k_{x},$ and due to the continuity of $%
\varepsilon _{\mathbf{k}}$even at the Dirac points. Therefore one is left
with the linear response.


\begin{figure}[ptb]\begin{center}
\centerline{\epsfxsize=7cm \epsffile{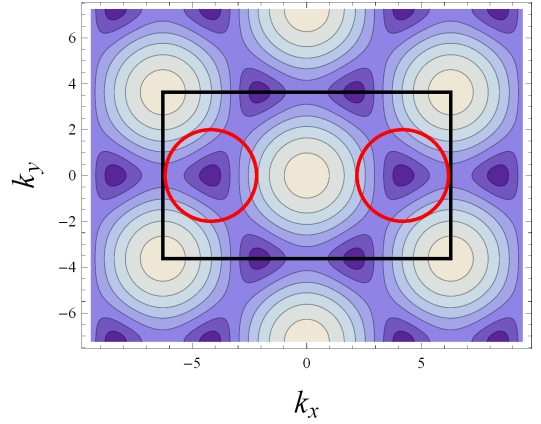} }
\caption{Fig.1 Constant energy map of the honeycomb-lattice Brillouin zone.
The rectangle outlines two Brillouin zones, the area of integration in Eqs.(%
\protect\ref{Jy_def} and \protect\ref{sigma_div1}). The circles show the
vicinity of the two Dirac points with radius $\Lambda $ for the integration
in Eq.(\protect\ref{sig_Diracpoint}).}\label{Fig1}%
\end{center}
\end{figure}

The conductivity can be divided into an apparently linearly divergent part
and a finite part, $\sigma \left( t\right) =\sigma _{BZ}\left( t\right)
+\sigma _{DP}\left( t\right) $. As will be explained shortly, the
contributions to the first term at large $t$ stem mostly from the immediate
neighbourhoods of the two Dirac points, while contributions to the second
come from whole Brillouin zone. The part diverging linearly with time can be
written (after some algebra) as a full derivative with respect to $k_{y}$ :

\begin{eqnarray}
\sigma _{BZ}\left( t\right) &=&t\sum_{\mathbf{k}}\frac{\left( h_{\mathbf{k}%
}h_{\mathbf{k}}^{\ast ^{\prime }}-h_{\mathbf{k}}^{\ast }h_{\mathbf{k}%
}^{\prime }\right) ^{2}-2\varepsilon _{\mathbf{k}}^{2}\left( h_{\mathbf{k}%
}h_{\mathbf{k}}^{\ast ^{\prime \prime }}+h_{\mathbf{k}}^{\ast }h_{\mathbf{k}%
}^{\prime \prime }\right) }{\varepsilon _{\mathbf{k}}^{3}}  \notag \\
&=&-4t\sum_{\mathbf{k}}\varepsilon _{\mathbf{k}}^{\prime \prime }.
\label{sigma_div1}
\end{eqnarray}%
This too vanishes upon integration over the Brillouin zone, since it is
again an integral of a derivative of a periodic function, albeit this
cancellation is nontrivial. In Fig. 2a the integrand of Eq.(\ref{sigma_div1}%
) is plotted within the integration area specified above, Eq.(\ref{integral}%
).The integrand is negative along the line connecting two Dirac points along
the field direction, for example

\begin{equation}
\mathbf{K}_{L}=2\pi \left( \frac{1}{3},\frac{1}{\sqrt{3}}\right) ,\ \ \ \ 
\mathbf{K}_{R}=2\pi \left( -\frac{1}{3},\frac{1}{\sqrt{3}}\right)  \label{K}
\end{equation}%
(recall that $a=1$ in our units), and positive elsewhere. In Fig. 2b several
cross sections for various $k_{x}$ values are shown. One observes, that as $%
k_{x}$ approaches $2\pi /3$ the integrand goes to $-\infty $ at $k_{y}=2\pi /%
\sqrt{3},$ corresponding to Dirac point $\mathbf{K}_{L}$. Consequently for
all $k_{x}\neq 2\pi /3,$the integrand is a regular function, and thus the
integral over $k_{y}$ vanishes. At $k_{x}=2\pi /3$ the integral over $k_{y}$
is finite, yet does not influence the two-dimensional integral.

\begin{figure}[ptb]\begin{center}
\centerline{\epsfxsize=7cm \epsffile{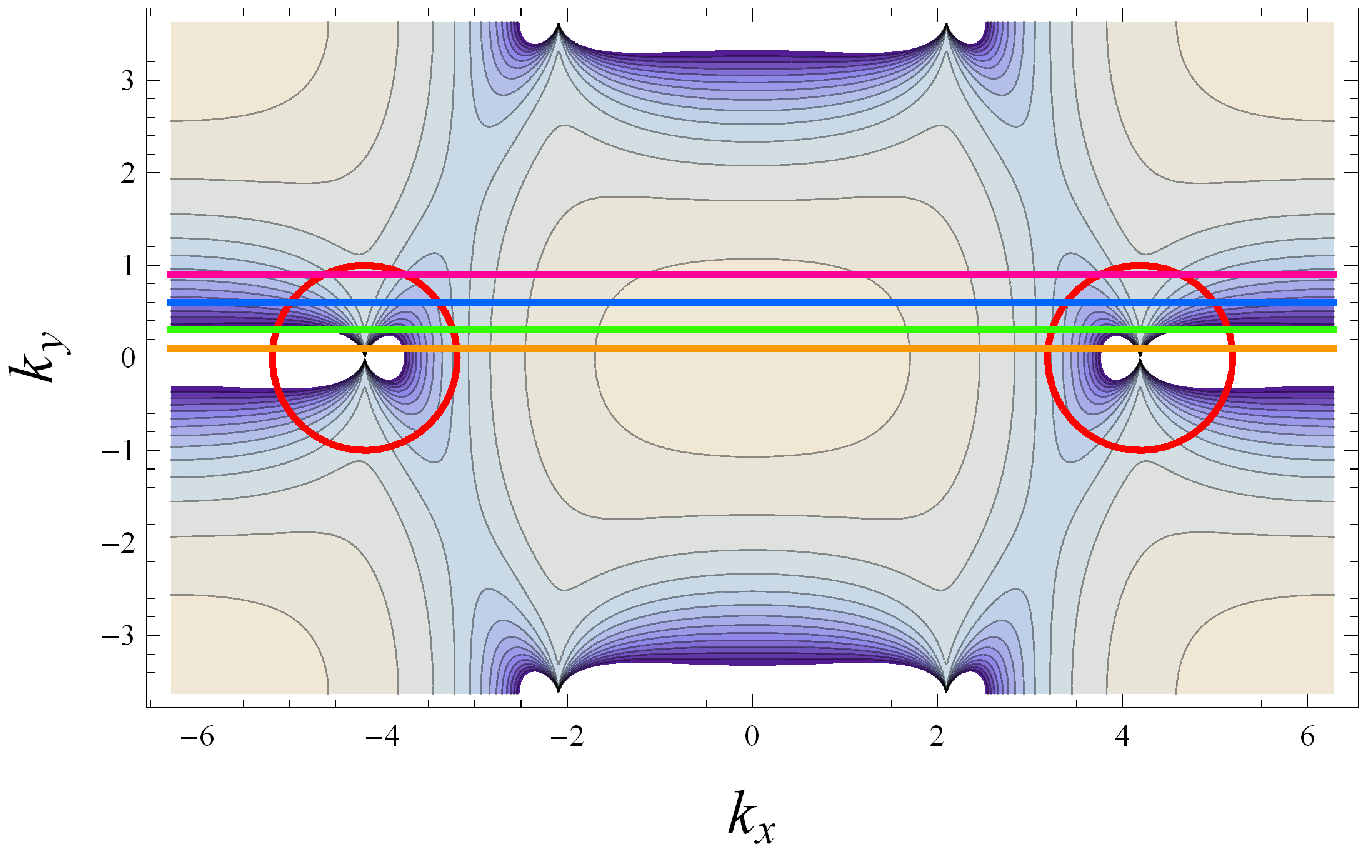} }
\caption{Fig. 2a. The integrand of $\protect\sigma _{BZ}$ Eq.(\protect\ref%
{sigma_div1}) is plotted within the integration area shown in Fig. 1.The
integrand is negative along the line connecting two Dirac points along the
field direction, and positive elsewhere.}\label{Fig2a}%
\end{center}
\end{figure}%

\bigskip

\begin{figure}[ptb]\begin{center}
\includegraphics[
natheight=2.7691in, natwidth=4.1632in, height=2.2381in, width=3.3486in]
{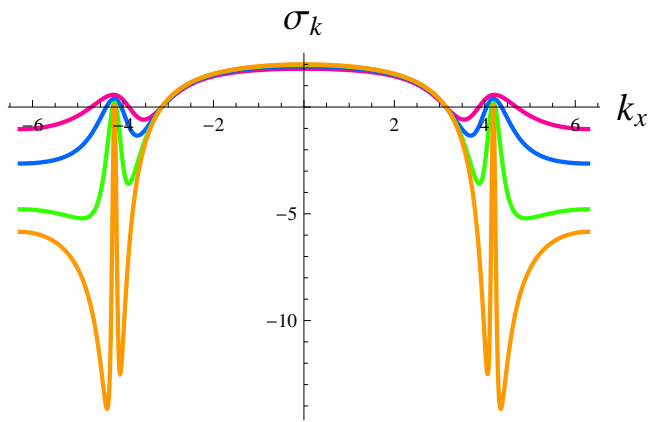}%
\caption{Fig. 2b. Several cross sections of the integrand of $\protect\sigma %
_{BZ}$ Eq.(\protect\ref{sigma_div1}) are shown for various $k_{x}$ values.}
\label{Fig2b}%
\end{center}\end{figure}%

The remaining part%
\begin{equation}
\sigma _{DP}\left( t\right) =-2\sum_{\mathbf{k}}\left( I_{\mathbf{k}%
}^{-}\right) ^{2}\sin \left( 2\varepsilon _{\mathbf{k}}t\right) \text{,}
\label{sig_k}
\end{equation}%
gives a finite result. Unlike the divergent part of the conductivity, the
integral for $t>>1$ (in units of $t_{\gamma }$) is dominated by
contributions from the vicinity of the two Dirac points (circles in Fig.2a
of radius $\Lambda \simeq 1$ (in units of $\hbar /a$)). Indeed the prefactor 
$\left( I_{\mathbf{k}}^{-}\right) ^{2}$ is bounded from above by $\Lambda
^{-2}$ and integral over the area outside the circles vanishes at $%
t>>1/\Lambda $. The contribution of a single Dirac point is obtained by
integrating to infinity (in polar coordinates centered at the Dirac point), 
\begin{eqnarray}
\frac{\sigma _{DP}}{2} &=&\frac{1}{\left( 2\pi \right) ^{2}}\int_{\varphi
=0}^{2\pi }\int_{q=0}^{\Lambda }\sin \left( \varphi \right) ^{2}\frac{\sin
\left( 2v_{g}qt\right) }{q}  \label{sig_Diracpoint} \\
&=&\frac{1}{4\pi }\int_{q^{\prime }=0}^{2v_{g}\Lambda t}\frac{\sin \left(
q^{\prime }\right) }{q^{\prime }}\text{.}  \notag
\end{eqnarray}%
At long time the upper limit can be replaced by infinity which results 
\begin{equation}
\sigma _{DP}=\frac{1}{2\pi }\func{Si}\left( \infty \right) =\frac{1}{4}\text{%
.}  \label{sig_DP}
\end{equation}%
Here one can safely multiply the result by the valley degeneracy $2$.
Returning to physical units, one obtains $\sigma =\sigma _{2}$, Eq.(\ref%
{sigma2}).

The dependence on time was calculated numerically. After an initial increase
on the natural time scale $t_{\gamma }\equiv \hslash /\gamma $ the
conductivity approaches $\sigma _{2}$ via oscillations, see Fig. 1 in ref. 
\cite{Lewkowicz}. The amplitude of oscillations decays as a power $\frac{%
\sigma }{\sigma _{2}}\approx 1+\frac{\sin \left( 2t\right) }{2t}$ (for $%
t>>t_{\gamma }$).

A physical picture of this resistivity without dissipation is as follows.
The electric field creates electron - hole excitations in the vicinity of
the Dirac points in which electrons behave as massless relativistic fermions
with the graphene velocity $v_{g}$ playing the role of the light velocity.
For such particles the absolute value of the velocity is $v_{g}$ and cannot
be altered by the electric field and is not related to the wave vector $%
\mathbf{k}$\textbf{. }On the other hand, the orientation of the velocity is
influenced by the applied field.\textbf{\ }The electric current is $e\mathbf{%
v}$, thus depending on orientation, so that its projection on the field
direction $y$ is increased by the field. An important observation, made for
example in ref.\cite{Sachdev}, is that the electric field, while creating
charge transport, does not change the overall momentum. As a consequence the
effects of impurities do not affect the pair creation in the same way they
affect "free" carriers. These two sources of current, namely creation and
reorientation are roughly of the same size in the linear response. As we
will see below, the situation changes at ballistic times when the linear
response breaks down.

\subsection{Pair creation rate}

In analogy to the linear response in the current, the pair creation rate per
unit area, as defined in Eq.(\ref{N_unren}), can be calculated and to
leading order in $\mathcal{E}$\ is: 
\begin{equation}
\frac{d}{dt}N_{0}=-2t\sum_{\mathbf{k}}\left[ I_{\mathbf{k}}^{-}\sin \left(
2\varepsilon _{\mathbf{k}}t\right) \right] ^{2}\mathcal{E}^{2}\text{.}
\label{N_pert}
\end{equation}%
The result of the numerical integration over the Brillouin zone was given in
Fig. 2 of ref. \cite{Rosenstein10}. It is dominated by the two Dirac points
and at large times (compared to $t_{\gamma }$) behaves as%
\begin{equation}
\frac{d}{dt}N_{0}\simeq \frac{2}{\pi }\mathcal{E}^{2}t\log \left( t\right) 
\text{.}  \label{N_pert1}
\end{equation}%
It is however well known that a dc electric field in QED (even with massive
relativistic electrons) renders the vacuum unstable \cite{Schwinger,Gitman}%
,\ with the "renormalized" number of pairs depending on $\mathcal{E}^{3/2}$.
Therefore nonperturbative effects are expected and will be studied below .

The notion of "renormalized" number of pairs is a consequence of the fact
that for unstable systems the Fermi level should be continuously
"renormalized" as function of time. In the first quantized formalism this
corresponds to a continuous modification of the "initial" state $v_{\mathbf{k%
}}\rightarrow v_{\mathbf{p}}=v_{\mathbf{k-}\mathcal{E}t}$ defining holes.
This leads to the following definition of the pair density \cite{Nussinov},

\begin{equation}
N_{p}\left( t\right) =2\sum_{\mathbf{k\in }BZ}\left\vert \left\langle \psi
\left( t\right) |v_{\mathbf{p}}\right\rangle \right\vert
^{2}=2\sum_{BZ}\left\vert \psi _{1}^{\ast }+\frac{h_{\mathbf{p}}^{\ast }}{%
\varepsilon _{\mathbf{p}}}\psi _{2}^{\ast }\right\vert ^{2}\text{.}
\label{N_ren}
\end{equation}%
It is used in the framework of the relativistic Dirac model and was recently
extensively discussed in ref.\cite{Cohen} in connection with graphene (using
the instanton approximation).

\subsection{ ac field}

A similar calculation within the dynamic first quantization formalism for
the evolution of the current density can be performed for an arbitrary time
dependence of the homogeneous electric field. Let us consider an arbitrary
time dependence of the $y$ component of the vector potential $A_{y}=-%
\mathcal{E}a\left( t\right) $, subject to a "switching on" condition, $%
a\left( t\right) =0$, for $t<0$. Fourier transforms are defined by%
\begin{equation}
a\left( t\right) =\int_{\omega }e^{-i\omega t}a\left( \omega \right) ;\text{
\ }\xi _{\mathbf{k}}\left( t\right) =\int_{\omega }e^{-i\omega t}\xi _{%
\mathbf{k}}\left( t\right) ,  \label{a}
\end{equation}%
and similarly for the current density and other quantities.$\ $The next to
leading order in $\mathcal{E}$\ first quantized tight binding equations are:

\begin{equation}
-a\left( \omega \right) 
\begin{pmatrix}
0 & h_{\mathbf{k}}^{\prime } \\ 
h_{\mathbf{k}}^{\ast \prime } & 0%
\end{pmatrix}%
u_{\mathbf{k}}+\left( 
\begin{array}{cc}
\varepsilon _{\mathbf{k}}-\omega & h_{\mathbf{k}} \\ 
h_{\mathbf{k}}^{\ast } & \varepsilon _{\mathbf{k}}-\omega%
\end{array}%
\right) \xi _{\mathbf{k}}\left( \omega \right) =0\text{.}  \label{corrAC}
\end{equation}%
The switching on condition, $\xi _{\mathbf{k}}\left( t<0\right) =0$, can be
taken into account by the $\omega +i0^{+}$ substitution regularizing the way
the Fourier transform is defined for zero frequency. From Eq.(\ref{corrAC})
one obtains

\begin{equation}
\xi _{\mathbf{k}}\left( \omega \right) =-\frac{a\left( \omega \right) }{%
\sqrt{2}\omega \left( 2\varepsilon -\omega \right) }\left( 
\begin{array}{c}
\omega h^{\ast }h^{\prime }/\varepsilon -h^{\ast }h^{\prime }-h^{\ast \prime
}h \\ 
-\omega h^{\ast \prime }+\varepsilon h^{\ast \prime }+h^{\ast 2}h^{\prime
}/\varepsilon%
\end{array}%
\right) .  \label{ksi_AC}
\end{equation}%
In the same order the ac conductivity is an integral over the sum of the
"paramagnetic" and the "diamagnetic" contributions given by

\begin{equation}
\frac{j_{\mathbf{k}}\left( \omega \right) }{a\left( \omega \right) }=4\frac{%
\left( h^{\ast }h^{\prime }-h^{\ast \prime }h\right) ^{2}}{\varepsilon
\left( 2\varepsilon +\omega \right) \left( 2\varepsilon -\omega \right) }-2%
\frac{h^{\ast }h^{\prime \prime }+hh^{\ast \prime \prime }}{\varepsilon }%
\text{.}  \label{sigma_AC}
\end{equation}%
Subtracting a full derivative (independent of frequency), one can rewrite
this as 
\begin{equation}
\frac{j_{\mathbf{k}}\left( \omega \right) }{a\left( \omega \right) }=\frac{%
\left( h^{\ast }h^{\prime }-h^{\ast \prime }h\right) ^{2}}{\varepsilon ^{3}}%
\frac{\omega ^{2}}{4\varepsilon ^{2}-\omega ^{2}}.  \label{sigAC}
\end{equation}

Real and imaginary parts, taking the $\omega +i0^{+}$ definition into
account, are

\begin{eqnarray}
\func{Re}\sigma &=&\frac{4}{i\omega }\sum_{\mathbf{k}}\frac{\left( h^{\ast
}h^{\prime }-h^{\ast \prime }h\right) ^{2}}{\varepsilon }  \notag \\
&&\times \lim_{s\rightarrow 0}\frac{4\varepsilon ^{2}-\omega ^{2}+s^{2}}{%
\left[ \left( 2\varepsilon +\omega \right) ^{2}+s^{2}\right] \left[ \left(
2\varepsilon -\omega \right) ^{2}+s^{2}\right] }  \notag \\
&=&\frac{1}{4}; \\
\func{Im}\sigma &=&\frac{4}{i\omega }\sum_{\mathbf{k}}\frac{\left( h^{\ast
}h^{\prime }-h^{\ast \prime }h\right) ^{2}}{\varepsilon }  \notag \\
&&\times \lim_{s\rightarrow 0}\frac{2\omega s}{\left[ \left( 2\varepsilon
+\omega \right) ^{2}+s^{2}\right] \left[ \left( 2\varepsilon -\omega \right)
^{2}+s^{2}\right] }  \notag \\
&=&0.  \notag
\end{eqnarray}%
Therefore ac and dc conductivity are equal, as was also shown in recent
calculations\cite{MacDonald}. This is consistent with both the Kubo formula
derivations \cite{Varlamov07} and optical experiments \cite{GeimScience08}.
Similar results were obtained in multiple layered graphene \cite{MacDonald}.

A similar calculation can be performed directly in the $t$ space (similar to
the dc case) and shows that after a short transient one obtains the $\sigma
_{2}$ value for the ac conductivity independent of frequency.

\section{IV. Beyond linear response. Weyl fermions, parity anomaly and the
Nielsen - Ninomiya theorem}

As will be discussed below, an electric field should not necessarily\ be
very small in graphene experiments, so that corrections to linear response
are of interest. In addition it is important to determine at what ballistic
time scale perturbation theory breaks down. In this section we present a
perturbative calculation of the leading nonlinear effect in both dc and ac,
and discuss the ways to regularize the Weyl model "correctly" in order to
calculate these corrections.

\subsection{Ultrarelativistic fermions near Dirac points}

The tight binding model employed here has two Dirac points around which the
spectrum becomes ultra - relativistic $\varepsilon =v_{g}\left\vert \Delta
_{L,R}\mathbf{k}\right\vert $, $\Delta _{L,R}\mathbf{k=k-K}_{L,R}$, see
Fig.1. In our units the graphene velocity is $v_{g}=\frac{\sqrt{3}}{2}$. The
matrix element of the tight binding Hamiltonian can be expanded around $%
K_{L} $ as

\begin{equation}
h_{\mathbf{k}}^{L}=v_{g}\exp \left( -i\frac{\pi }{3}\right) \left( \Delta
k_{x}+i\Delta k_{y}\right) \text{.}  \label{Weylm}
\end{equation}%
The Weyl field describing the "left handed"\cite{Weinberg,Smit} fermions $%
\psi ^{L}$, namely a field satisfying the relativistic Dirac equation with
zero fermion mass%
\begin{eqnarray}
i\partial _{t}\psi _{1}^{L} &=&v_{g}\left( \partial _{x}+i\partial
_{y}\right) \psi _{2}^{L}  \label{eqWeyl} \\
i\partial _{t}\psi _{2}^{L} &=&v_{g}\left( \partial _{x}-i\partial
_{y}\right) \psi _{1}^{L}\text{,}  \notag
\end{eqnarray}%
can be constructed by the unitary transformation 
\begin{equation}
\psi _{1}=\psi _{1}^{L},\ \psi _{2}=e^{-i\frac{\pi }{3}}\psi _{2}^{L}\text{.}
\label{unitary}
\end{equation}%
The matrix element around the second Dirac point $\mathbf{K}_{R}$ is
different, 
\begin{equation}
h_{\mathbf{k}}^{R}=v_{g}\left( \Delta k_{x}-i\Delta k_{y}\right)  \label{hWp}
\end{equation}%
and the Weyl fermions are "right handed" particles that obey%
\begin{eqnarray}
i\partial _{t}\psi _{1}^{R} &=&v_{g}\left( \partial _{x}-i\partial
_{y}\right) \psi _{2}^{R} \\
i\partial _{t}\psi _{2}^{R} &=&v_{g}\left( \partial _{x}+i\partial
_{y}\right) \psi _{1}^{R},  \notag
\end{eqnarray}%
without any unitary transformation required. They belong to a different
representation of the 2+1 dimensional Lorentz group \cite{Luscher}.

This is not just a peculiarity of the model, but a rather general feature 
\cite{GusyninPRB06} of massless fermions on a lattice \cite{Smit,Nielsen}.
It is well known that in order to get a massless spectrum of fermionic
excitations with any ultraviolet cutoff (hexagonal lattice is an example),
they come in multiple locations on the Brillouin zone (species "doubling").
In the Hamiltonian formulation the multiplicity is $2^{D}$, where $D$ is the
dimensionality of space, $D=2$ in graphene. In addition, the graphene
fermions are "staggered", meaning the spinor components "live" on different
sublattices of the honeycomb lattice. This reduces the doubling to $%
2^{D-1}=2 $. The doubling is intimately linked to the parity (discrete
chiral symmetry) \cite{Luscher, Smit}. The two Dirac points have opposite
chirality so that there is no "parity anomaly".

It is sometimes claimed in condensed matter literature that, at least while
doing linear response, one can concentrate\ on the two neighbourhoods of the
Dirac points and neglect the rest of the Brillouin zone. Even more, often
the calculation's result is just multiplied by the factor $2$ (the valley
degeneracy). Below we show that this is generally not an accurate
description of what happens. This is not an academic question, since the low
energy Weyl theory is simpler than the tight binding model and will be used
in the next section to extend the calculations into higher orders in the
electric field. One therefore needs a proper ultraviolet regularization of
the Weyl model. Similar regularization issues are well known in field theory
whenever chiral anomaly is encountered. Roughly, a "correct" regularization
should respect the chiral symmetry leading to important constraints on the
number and charges of massless fermions. Otherwise the model is called
"anomalous" and results of perturbative calculations become arbitrary. The
most famous example is the requirement that in each generation of elementary
particles (leptons and quarks) the sum of charges should be zero \cite%
{Peskin}.

We therefore discuss in some detail the cancellation of infrared
divergencies and the correct application of the "ultra - relativistic"
approximation to the tight binding model.

\subsection{Cancellation of ultraviolet divergencies and the approximate
chiral symmetry.}

The ultra - relativistic approximation, Eq.(\ref{Weylm}), fails when applied
to the first term in the expression for conductivity $\sigma _{BZ}$ given in
Eq.(\ref{sigma_div1}). At first glance the integral in Eq.(\ref{sigma_div1})
is dominated by the two Dirac points since the integrand diverges there, see
Fig. 2. For the both (widely separated) regions of the Brillouin zone, $%
K_{L} $ and $K_{R}$, it has the same asymptotic form%
\begin{equation}
\frac{\left( hh^{\ast ^{\prime }}-h^{\ast }h^{\prime }\right) ^{2}}{%
2\varepsilon ^{3}}-\frac{hh^{\ast ^{\prime \prime }}+h^{\ast }h^{\prime
\prime }}{\varepsilon }\approx -\sqrt{3}\frac{\left( \triangle k_{x}\right)
^{2}}{\left\vert \triangle \mathbf{k}\right\vert ^{3}}\text{.}
\label{integrand+Dirac}
\end{equation}%
The integral over the neighbourhood of each Dirac point converges in the
"infrared" (here meaning $\mathbf{k}\rightarrow \mathbf{K}_{L,R}$) due to
the Jacobian $\left\vert \triangle \mathbf{k}\right\vert $, but is linearly
divergent in the "ultraviolet" and both have the same sign, see Fig.2. Hence
the integration cannot be extended to infinity and the size of the Brillouin
zone serves as a natural ultraviolet cutoff.

It is important to note that there is no cancellation of the divergence
between the Dirac points since both have the same sign! The divergencies are
however canceled by the contributions from regions of the Brillouin zone
between the Dirac points in which the ultra - relativistic approximation is
not valid. Therefore in the "ohmic" regime during ballistic times one is not
allowed to neglect states far from the Dirac points and replacement by the
Weyl equation is incorrect. Due to cancellation of the whole "divergent"
term, Eq.(\ref{sigma_div1}), one can devise an appropriate regularization in
the UV in which these contributions are cancelled and even generalize the
procedure to higher orders in the electric field. We propose and use such a
scheme below in section IVC. Simple recipes like the momentum cutoff
regularization (circles in Fig.1) or giving the fermions an infinitesimal
mass \cite{GusyninPRB06}, commonly used, may lead to unphysical results. As
a consequence more sophisticated regularizations like the $\zeta $-function
regularization \cite{Beneventano09}, "dimensional regularization" \cite%
{Peskin} and "Pauli - Villars" were developed for continuous field theories
like the Weyl model. The tight binding model is very similar in this respect
to the "lattice Hamiltonian" regularizations of the field theory
(Hamiltonian meant here with time kept continuous) and also satisfies the
chiral invariance criterion \cite{Smit}.

\subsection{Nonlinear response in dc. Where does the linear response break
down?}

Since the current density is an odd function of the electric field, the
leading nonlinear correction to it appears in the third order in the field.
The calculation along the lines of subsection IIIA is quite straightforward
albeit tedious. One can use the Weyl model instead of the tight binding
model due to the following reason. It is well known in field theory that
generally chiral anomaly effects, including the ultraviolet divergencies
discussed in section IIID, appear only in one loop calculations \cite{Peskin}%
. We checked and found that indeed up to the third order the expression is
finite in the ultraviolet. Within the dynamical approach it is natural to
perform the calculations without Fourier transforming into the $\omega $
space.

Up to the third order the current density is 
\begin{equation}
j\left( t\right) =\frac{\mathcal{E}}{4}\left[ 1+\frac{3}{64}t^{4}\mathcal{E}%
^{2}+O\left( \mathcal{E}^{4}\right) \right] \text{.}  \label{J_correction}
\end{equation}%
The correction therefore is growing as a large power of the ballistic time.
It becomes as large as the leading order for $t=2^{3/2}3^{-1/4}\mathcal{E}%
^{-1/2}t_{\gamma }$. Hence the perturbation theory breaks down on the time
scale of 
\begin{equation}
t_{nl}=\mathcal{E}^{-1/2}t_{\gamma }\text{.}  \label{tnl_pert}
\end{equation}%
As will be seen in Section V, this agrees well with the crossover time
obtained from a nonperturbative calculation. Of course this time should be
larger than any other possible relaxation time (due to impurities, phonons
etc.) present in the system. A similar expansion was obtained for the number
of electron-hole pairs \cite{Rosenstein10}.

The result is the same for the tight binding and the corresponding Weyl
model. The only issue would be the first order calculation within the Weyl
scheme since it is divergent in the ultraviolet. In Weyl model we used the
momentum cutoff that will be described in section IVB.

\subsection{ac response and the third harmonic generation}

An analogous calculation for the ac field $E=\mathcal{E}\cos \left( \omega
t\right) $ switched on at $t=0$ results in 
\begin{eqnarray}
j\left( t\right) /\mathcal{E} &=&-\frac{1}{8\omega ^{4}}\mathcal{E}^{2}+%
\func{Re}\sigma \cos \left( \omega t\right) +\func{Im}\sigma \sin \left(
\omega t\right) +\sigma _{3\omega }\cos \left( 3\omega t\right)  \notag \\
&&+O\left( \mathcal{E}^{4}\right) .
\end{eqnarray}%
The first term is just the reflection of the initial conditions and is non -
universal. In the limit $\omega \rightarrow 0$ one recovers the dc result.
The corrected value of the real (dissipative) part of the ac conductivity is

\begin{equation}
\func{Re}\sigma =\frac{1}{4}\left[ 1+\left( \frac{63}{128\omega ^{4}}-\frac{9%
}{64\omega ^{2}}t^{2}\right) \mathcal{E}^{2}\right] .  \label{Resigma}
\end{equation}%
One observes that the perturbation theory is inapplicable for 
\begin{equation}
\omega ^{2}<\mathcal{E}\text{ \ and \ }t>\omega /\mathcal{E}\text{.}
\label{conditions}
\end{equation}%
This is less restrictive compared to the dc condition, Eq.(\ref{tnl_pert})
for $\omega \mathcal{>E}^{1/2}t_{\gamma }^{-1}$.

The present formulas can be compared with the results of the dynamical
calculation beyond linear response by Mishchenko \cite{MishchenkoPRL09} in
which a phenomenological model of relaxation was employed. For the inverse
relaxation time $\Gamma $, his nonperturbative result is%
\begin{eqnarray*}
\func{Re}\sigma &=&\frac{\omega ^{2}\Gamma ^{2}}{2v_{g}^{2}\mathcal{E}^{2}}%
\left[ \left( 1+\frac{v_{g}^{2}\mathcal{E}^{2}}{\omega ^{2}\Gamma ^{2}}%
\right) ^{1/2}-1\right] \\
&\approx &\frac{1}{4}\left[ 1-\frac{3\mathcal{E}^{2}}{16\omega ^{2}\Gamma
^{2}}+O\left( \mathcal{E}^{4}\right) \right] \text{.}
\end{eqnarray*}%
This is consistent with the second correction terms in Eq.(\ref{Resigma}).

The inductive part was absent in linear response and now appears for the
first time:%
\begin{equation}
\func{Im}\sigma =\frac{3^{3}}{2^{8}\omega ^{3}}t\mathcal{E}^{2}\text{.}
\label{Imsigma}
\end{equation}%
Due to the two conditions, Eq.(\ref{conditions}), it is much smaller than $%
\sigma _{2}$. The third harmonic is generated with the real part%
\begin{equation*}
\sigma _{3\omega }=\frac{1}{2^{9}\omega ^{4}}\mathcal{E}^{2}\text{,}
\end{equation*}%
while the inductive part is absent at the present order.

\section{V. The exact solution of the first quantized tight binding equations%
}

\subsection{Application of the Floquet theory and reduction to 1D}

It is a peculiar property of the tight binding matrix Eq.(\ref{H_matrix})
that the solution for arbitrary $k_{y}$ can be reduced to that for $k_{y}=0$%
. Shifting the time variable to $\overline{t}=t-k_{y}/\mathcal{E}$, one can
define an 'universal wave function' 
\begin{equation}
\ \overline{\psi }\left( \overline{t}\right) =\psi \left( t\right) .
\label{psibar}
\end{equation}%
The Schroedinger equation (\ref{Schroedinger1})\ is now void of any $k_{y}$
dependence,

\begin{eqnarray}
i\partial _{\overline{t}}\overline{\psi }_{1} &=&-\left( e^{-2i\Omega 
\overline{t}}+be^{i\Omega \overline{t}}\right) \overline{\psi }_{2}
\label{Schroedpsibar} \\
i\partial _{\overline{t}}\overline{\psi }_{2} &=&-\left( e^{2i\Omega 
\overline{t}}+be^{-i\Omega \overline{t}}\right) \overline{\psi }_{1}\text{,}
\notag
\end{eqnarray}%
where the dimensionless frequency $\Omega \equiv \mathcal{E}/\left( 2\sqrt{3}%
\right) $ is in units of \ $t_{\gamma }^{-1}$. The $k_{y}$ component of the
momentum enters\ the solution via the initial conditions Eq.(\ref{u}) only.
In terms of the universal function it takes the form 
\begin{equation}
\psi _{\mathbf{k}}\left( t=0\right) =\overline{\psi }\left( \overline{t}%
=-k_{y}/\mathcal{E}\right) =u_{\mathbf{k}}\text{.}  \label{initial_bar}
\end{equation}

Taking the Fourier transform of Eqs. (\ref{Schroedpsibar}), the equations in
frequency space turn into recursion relations:

\begin{eqnarray}
\omega \overline{\psi }_{1}\left( \omega \right) &=&-\overline{\psi }%
_{2}\left( \omega -2\Omega \right) -b\overline{\psi }_{2}\left( \omega
+\Omega \right)  \label{eq1wbar} \\
\omega \overline{\psi }_{2}\left( \omega \right) &=&-\overline{\psi }%
_{1}\left( \omega +2\Omega \right) -b\overline{\psi }_{1}\left( \omega
-\Omega \right) \text{.}  \label{eq2wbar}
\end{eqnarray}%
Floquet theory (and the recursion relations) assure that the frequencies are
discrete and form two series both indexed by an integer $m$ 
\begin{equation}
\omega _{m}=\nu +3\Omega m\text{.}  \label{omegam}
\end{equation}%
The "central" frequency $\nu $ will take generally two incommensurate
values. Writing the Fourier amplitude $\overline{\psi }_{1}\left( \omega
_{m}\right) $ as $p_{m}$, one obtains, combining the two equations, Eq.(\ref%
{eq1wbar}) and Eq.(\ref{eq2wbar}), the forward and the backward recursions,

\begin{eqnarray}
p_{m} &=&\frac{\omega _{m}-2\Omega }{b}\left[ \omega _{m}-3\Omega -\frac{1}{%
\omega _{m}-5\Omega }-\frac{b^{2}}{\omega _{m}-2\Omega }\right] p_{m-1} 
\notag \\
&&-\frac{\omega _{m}-2\Omega }{\omega _{m}-5\Omega }p_{m-2},
\label{f_recursion1}
\end{eqnarray}%
\begin{eqnarray}
p_{m} &=&\frac{\omega _{m}+\Omega }{b}\left[ \omega _{m}+3\Omega -\frac{1}{%
\omega _{m}+\Omega }-\frac{b^{2}}{\omega +4\Omega }\right] p_{m+1}  \notag \\
&&-\frac{\omega _{m}+\Omega }{\omega _{m}+4\Omega }p_{m+2},
\label{b_recursion1}
\end{eqnarray}%
with normalization $p_{0}=1$. The $p_{1}$ and $\nu $ have to be determined
by the consistency between the forward and the backward equations.

Generally there are two independent Floquet frequencies, however our system
is special with the exact relation 
\begin{equation}
\nu ^{+}+\nu ^{-}=2\Omega ,  \label{omega_sum}
\end{equation}%
which can be proven rigorously. This greatly simplifies the solution. To
obtain a simple approximation, we expand around the easily solvable case of $%
k_{x}=\pi $ corresponding to $b=0$. In this case the solution consists of
two frequencies 
\begin{equation}
\nu ^{\pm \left( 0\right) }=\Omega \pm \sqrt{1+\Omega ^{2}},
\label{Floquet freq}
\end{equation}%
where the superscript $^{\left( 0\right) }$ denotes zeroth order in $b$. For
experimentally accessible cases $\Omega <<1$, the Floquet frequencies are
close to $\pm 1$. We expand the Fourier coefficients in powers of $b$: 
\begin{equation}
p_{m}=p_{m}^{\left( 0\right) }+bp_{m}^{\left( 1\right) }+b^{2}p_{m}^{\left(
2\right) }\,\text{,}  \label{pmexpansion}
\end{equation}%
where we do not distinguish for the time being between the two Floquet
branches. Let us look for a solution with the initial choice 
\begin{equation}
p_{m}^{\left( 0\right) }=p^{\left( 0\right) }\delta _{m}\text{,}  \label{p0}
\end{equation}%
which in particular means that $p_{1}=0$ at this order. It can be shown that
the "central" frequency $\nu ^{+}\,\ $does not have a first order correction
in $b$. Using this fact the forward recursion can be rewritten as: 
\begin{eqnarray}
&&\left( \omega _{m}+\Omega \right) \left( \omega _{m}^{2}-2\Omega \omega
_{m}-1\right) p_{m} \\
&=&b\left[ \left( \omega _{m}+\Omega \right) p_{m-1}+\left( \omega
_{m}-2\Omega \right) p_{m+1}\right] +b^{2}\left( \omega _{m}-2\Omega \right)
p_{m}\text{.}  \notag
\end{eqnarray}%
Inserting the expansions for $p_{m}$ and $\nu $ into this relation\ yields
the expression for $\nu ^{+\left( 2\right) }$,

\begin{eqnarray}
\nu ^{+\left( 2\right) } &=&\frac{\nu ^{\left( 0\right) }/2-\Omega }{\nu
^{\left( 0\right) 2}-\Omega ^{2}}-\frac{\nu ^{\left( 0\right) }-5\Omega }{%
6\Omega \left( \nu ^{\left( 0\right) }-\Omega \right) \left( \nu ^{\left(
0\right) }-2\Omega \right) \left( 2\nu ^{\left( 0\right) }-5\Omega \right) }
\notag \\
&&+\frac{\nu ^{\left( 0\right) }-2\Omega }{6\Omega \left( 2\nu ^{\left(
0\right) }+\Omega \right) \left( \nu ^{\left( 0\right) 2}-\Omega ^{2}\right) 
}\text{,}  \label{pert1}
\end{eqnarray}%
so that the positive Floquet frequency (up to second order) is $\nu ^{+}=\nu
^{\left( 0\right) }+b^{2}v^{\left( 2\right) }$. Moreover, it turns out that
the expansion in $b$ converges in the whole relevant range, $0<b\leq 2$, and
greatly assists the numerical solution of the recursion relation.

After the coefficients $p_{m}^{\pm }$ and Floquet frequencies $\nu ^{\pm }$
are found, the solution of Eq. (\ref{Schroedpsibar}) is written in terms of
the Fourier series:%
\begin{eqnarray}
\overline{\psi }_{\mathbf{k}}^{1}\left( \overline{t}\right) &=&\sum_{s=\pm
}A^{s}\sum_{m=-\infty }^{\infty }p_{m}^{s}e^{-i\omega _{m}^{s}\overline{t}};%
\text{ \ }  \label{Fourier} \\
\overline{\psi }_{\mathbf{k}}^{2}\left( \overline{t}\right) &=&\sum_{s=\pm
}A^{s}\sum_{m=-\infty }^{\infty }\frac{p_{m}^{s}+bp_{m-1}^{s}}{\omega
_{m}^{-s}}e^{i\omega _{m}^{-s}\overline{t}}\text{.}  \notag
\end{eqnarray}%
The second line follows from Eq.(\ref{eq2wbar}), $\omega _{m}^{s}$ are
defined in Eqs.(\ref{omegam}) and (\ref{pert}) and we utilized Eq.(\ref%
{omega_sum}) for simplification. The two complex coefficients $A^{s}$ are
fixed by the initial conditions Eq.(\ref{initial_bar}). This solution is
used to calculate the evolution of the current density, the energy and the
number of electron - hole pairs.

\subsection{Results. Nonlinear regime and Bloch oscillations.}

The current density divided by the electric field (in physical units), $%
\sigma \left( t\right) \equiv J_{y}\left( t\right) /E$, is shown in Fig. 3
in ref. \cite{Rosenstein10} for various values of the dimensionless electric
field $\mathcal{E}=E/E_{0}$. The (microscopic) unit of the electric field is
very large $E_{0}=\frac{\gamma }{ea}=10^{10}V/m$, so that at realistic
fields $\mathcal{E}<<1$. After an initial fast increase on the microscopic
time scale $t_{\gamma }$, $\sigma \left( t\right) $ approaches the universal
value $\sigma _{2}=\frac{\pi }{2}\frac{e^{2}}{h}$ and settles there,
consistent with the linear response, calculated in Section IIIB. Beyond the
crossover time $t_{nl}$ the conductivity rises linearly above the constant
"universal" value $\sigma _{2}$: 
\begin{equation}
\frac{J\left( t\right) }{E}=\sigma _{2}\lambda \mathcal{E}^{1/2}\frac{t}{%
t_{\gamma }},  \label{J2}
\end{equation}%
where $\lambda =2^{5/2}3^{1/4}\pi ^{-2}\simeq 0.75$. This is consistent with
the Landau - Zener calculation (instantons) within the Weyl model \cite%
{GavrilovPRD96,Dora10}, see below. The crossover time, defined by $J\left(
t_{nl}\right) /E=\sigma _{2}$, is therefore 
\begin{equation}
t_{nl}=\lambda ^{-1}\mathcal{E}^{-1/2}t_{\gamma }\text{,}  \label{t_cros}
\end{equation}%
and is consistent with the perturbation theory breakdown ballistic time, Eq.(%
\ref{tnl_pert}). This linear increase regime can be considered as a
precursor of the Bloch oscillations, but is still universal in the sense
that it depends solely on neighborhood of the Dirac points.

\begin{figure}[ptb]\begin{center}
\includegraphics[
natheight=4.3362in, natwidth=6.5812in, height=2.2364in, width=3.3806in]
{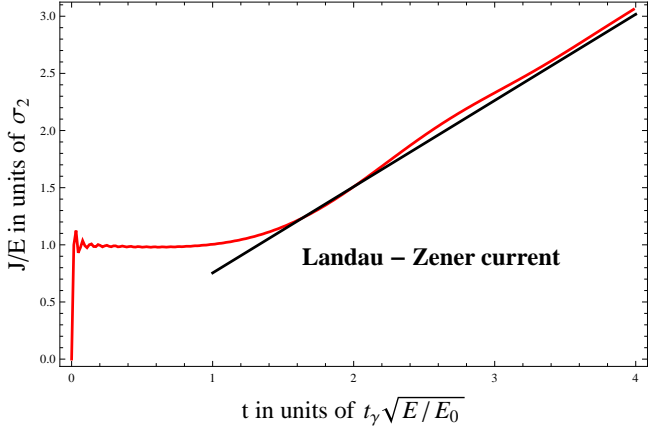}%
\caption{Fig. 3. The time evolution of the dc conductivity in units of $%
\protect\sigma _{2}.$The time is scaled with $\mathcal{E}^{-1/2}.$The
simulations were performed in the field range $\mathcal{E}=2^{-12}-2^{-9}$
with UV cutoff equal to $1/a$. The straight line is the asymptotic linear
behaviour given by Eq.(\protect\ref{J2}).}\label{Fig3}%
\end{center}\end{figure}%

The current on the time scale of order

\begin{equation}
t_{B}=\frac{8\pi }{\sqrt{3}}\mathcal{E}^{-1}t_{\gamma }=4\pi \frac{\gamma }{%
eEv_{g}}\text{,}  \label{t_Bloch}
\end{equation}%
exhibits Bloch oscillations, as is seen in Fig.4 of ref. \cite{Rosenstein10}%
, where larger fields in the range $\mathcal{E}=2^{-8}-2^{-5}$ are shown. It
turns out that the current vanishes at points given \textit{exactly }at
multiples of $t_{B}/2$ with $t_{B}$ being the period of the Bloch
oscillations. One observes a peculiar feature that (apart from the
"relativistic" initial constant segment) the time dependence of $\sigma
\left( t\right) $ is similar for different electric fields. Indeed, if one
plots $J/\sqrt{E}$ versus $tE$, all the curves nearly coincide. Moreover,
one obtains the following excellent fit for the current 
\begin{equation}
\frac{J\left( t\right) }{E}=\sqrt{3}\sigma _{2}\mathcal{E}^{-1/2}\sin \left( 
\frac{\sqrt{3}t}{4t_{\gamma }}\mathcal{E}\right) \text{.}  \label{sin}
\end{equation}

\bigskip

\begin{figure}[ptb]\begin{center}
\includegraphics[
natheight=4.28in, natwidth=6.5812in, height=2.2364in, width=3.4229in]
{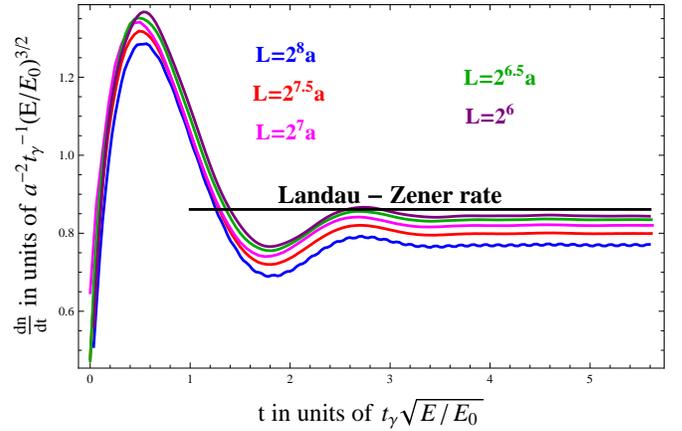}%
\caption{Fig. 4. The electron-hole pair creation rate as function of time in
units of $\mathcal{E}^{-1/2}t_{\protect\gamma }$for field $\mathcal{E}%
=2^{-10}$. The rate is scaled with $\mathcal{E}^{3/2}$. The ultraviolet
cutoff is always $1/a,$ while different curves represent different values
for the infrared cutoff $L/a$.}\label{Fig4}%
\end{center}\end{figure}%

The Bloch time is approximately the time required for the electric field to
shift the momentum across the Brillouin zone $\Delta p_{y}=eEt_{B}\sim \hbar
/a$. This time scale is very long for experimentally achieved fields, much
longer than the ballistic flight time. For example in a sample of submicron
length, $L=0.5\mu m$, $t_{bal}\simeq 2.3\cdot 10^{3}t_{\gamma }$. If one
assumes that the electric current can reach $I_{\max }=1mA$, so for a
typical width $W=1.5\mu m$ one has an electric field $E_{\max }=\frac{%
I_{\max }}{W\sigma _{2}}=10^{7}V/m\ $corresponding to $\mathcal{E}=10^{-3}$
(the voltage in such case would be quite large $V_{\max }=E_{\max }L=5V$).
The first maximum of the Bloch oscillation will be seen at flight time of $%
t_{B}/4=3.6\cdot 10^{3}t_{\gamma }$, which is of the same order as $t_{bal}$%
. If one uses a value of the current typical to transport measurements $%
I=50\mu A$, the electric field is just $E=5\cdot 10^{5}V/m\ $corresponding $%
\mathcal{E}=5\cdot 10^{-5}$ (voltage $V=250mV$), $t_{B}/4=7.2\cdot
10^{4}t_{\gamma }>>t_{bal}$ and is therefore out of reach. See, however a
recent proposal \cite{Dragoman08}.

\subsection{Crossover at $t_{nl}$ in the Weyl model}

It is expected that the transition to the nonlinear regime is dominated by
the neighborhoods of the Dirac points. Therefore it can be obtained also
within the Weyl model provided it is properly regularized in the ultraviolet
region consistent with chiral symmetry, as was discussed in section IV. For
the calculation of the current density it suffices to renormalize the
electric current by subtracting the UV divergent term Eq.(\ref%
{integrand+Dirac}):%
\begin{equation}
J_{y}=-\frac{4v_{g}}{\left( 2\pi \right) ^{2}}\int_{\left\vert \mathbf{k}%
\right\vert <\Lambda }\left[ \frac{1}{2}\left( \psi _{2}^{\ast }\psi
_{1}+\psi _{1}^{\ast }\psi _{2}\right) +\frac{k_{x}^{2}}{\left\vert \mathbf{k%
}\right\vert ^{3}}t\mathcal{E}\right] \text{.}  \label{Jy_reg}
\end{equation}

The results are practically indistinguishable compared to the tight binding
model for times larger than a microscopic time scale $t_{\gamma }$ and much
smaller than the Bloch time. Some examples of the tight binding model were
presented in Fig. 3 of ref. \cite{Rosenstein10}. In Fig. 3 the conductivity
in units of $\sigma _{2}$ is given as a function of time scaled with $%
\mathcal{E}^{-1/2}$. The function is the same for all electric fields as can
be seen from scaling properties of the Weyl equations. The simulations were
performed in the field range $\mathcal{E}=2^{-12}-2^{-9}$ with UV cutoff
equal to $1/a$. The straight line is the asymptotic linear behaviour given
by Eq.(\ref{J2}). The asymptotic form is attained soon after $t_{nl}$, so it
looks like quite a sharp crossover.

Therefore this model can be effectively used to study the transition to the
rapid creation of the electron - hole pairs and the creation of the electron
- hole plasma, but naturally cannot be used to see the transition to the
Bloch oscillation regime.

Below we discuss the nature of the crossover to the nonlinear behaviour and
possible new physics of the electron - hole plasma which might emerge.

\section{VI. Physical picture of the crossover to the nonlinear response}

\subsection{Where does the energy go?}

Beyond linear response one does not expect the current density to hold up to
its linear response value indefinitely. The situation differs from that in
diffusive systems in which energy dissipates into heat due to inelastic
scattering off impurities and phonons. In this case the system is not an
isolated one but becomes part of a larger system including a "background".
In an isolated ballistic system in a constant electric field the energy
initially increases, as follows from the following argument. The total
energy, Eq.(\ref{U_def}), of electrons can be written in the first quantized
formalism as 
\begin{equation}
U\left( t\right) =2\left\langle \psi \left( t\right) \left\vert H\right\vert
\psi \left( t\right) \right\rangle \text{.}  \label{energy}
\end{equation}%
It can be shown using Eq.(\ref{Schroedinger}), that the power in this driven
system is proportional to the current density

\begin{equation}
\frac{d}{dt}U=2\left\langle \psi \left\vert \frac{d}{dt}H\right\vert \psi
\right\rangle =-2eE\left\langle \psi \left\vert \frac{\partial H_{\mathbf{p}}%
}{\partial p_{y}}\right\vert \psi \right\rangle =EJ_{y}\left( t\right) \text{%
,}  \label{power}
\end{equation}%
like in dissipative systems. Consistently, as was shown above, the ballistic
system has a finite conductivity like a dissipative system. We will refer to
this as to "quasi - Ohmic" behaviour.

Since the model does not provide a channel of dissipation, where does this
"Joule - like" heat $\sigma E^{2}$ go? The dynamical approach allows us to
calculate the evolution of energy going beyond linear response. The energy
of the system (calculated in a way similar to the current) is increasing
continuously if no channel for dissipation is included. Therefore the
conductivity originates in creation of pairs near the Dirac points with an
additional contribution due to the turning of particles toward the field's
direction. To gain more insight into the nature of the crossover to
nonlinear response we also calculated the evolution of number of the
electron - hole pairs during the ballistic flight.

\subsection{Schwinger's pair creation formula and graphene}

The electron - hole creation rate by the dc electric field, $\frac{d}{dt}%
N_{p}$, with renormalized $N_{p}$ defined in Eq.(\ref{N_ren}), is shown in
Fig. 4 as function of time for field $\mathcal{E}=2^{-10}$. Such a field $%
E=2^{-10}E_{0}=10^{7}V/m$ is quite realistic \cite{SinghPRB09}. The rate is
scaled with $\mathcal{E}^{3/2}$, while the time is given in units of $%
\mathcal{E}^{-1/2}t_{\gamma }$. The ultraviolet cutoff is always $1/a,$
while different curves represent the infrared cutoff $L/a$. The length to
width aspect ratio, $L/W$, is taken to be $1$. The results do not depend
significantly on it for all values presented, as long as $1/4<L/W<4$.

The time dependence of the rate exhibits the same time scales as the current
density. At times smaller than $t_{nl}$ the perturbative formula, Eq.(\ref%
{N_pert}) is valid. Immediately after switching on the electric field (times
of order $t_{\gamma }$) the rate behaves as $t^{3}$. For $t_{\gamma
}<t<t_{nl}$ the pair creation rate per unit area rises linearly (with
logarithmic corrections) according to Eq.(\ref{N_pert1}). However it is
clear from Fig. 4 that the expansion breaks down at $t_{nl}$ when the rate
stabilizes, approaching the value (in our units of $a^{-2}t_{\gamma }^{-1})$

\begin{equation}
\frac{d}{dt}N_{p}=\frac{8}{\pi ^{2}v_{g}^{1/2}}\mathcal{E}^{3/2}.
\label{Schwinger}
\end{equation}

The power dependence of the rate on electric field, $E^{3/2}$ is the same as
the rate of the vacuum breakdown due to the pair production calculated
beyond perturbation theory by Schwinger in the context of particle physics
(when generalized to the 2+1 dimensions and zero fermion mass \cite%
{Schwinger,Cohen}). It is not surprising since the power $3/2$ is dictated
by the dimensionality, assuming the ultra - relativistic approximation is
valid. However the physical meaning is somewhat different.

Note that the definition of the (renormalized) particle number, see Eq.(\ref%
{N_ren}), is different from the classical Schwinger's path integral
definition (which actually determines the "vacuum decay" rate rater than the
pair creation rate). The two are not the same within the Weyl model, as was
shown asymptotically in the limit of large times in \cite{Nussinov}, and we
obtain their value. The calculation of the number of pairs can be
approximately performed using the instanton approach initiated by Nussinov
in the context of particle physics \cite{Nussinov} and extended in \cite%
{GavrilovPRD96,Cohen}. In condensed matter physics the method is known as
the Landau - Zener probability \cite{Dora10,Santoro}. It provides an
intuitive picture of Schwinger's pair creation rate. Unfortunately this
picture cannot be extended to ballistic times smaller than $t_{nl}$ in which
one cannot use the asymptotic large time expressions.

Note that in the Boltzmann equation approach to ballistic transport
developed in ref. \cite{Sachdev} the renormalized pair number was utilized.
This is connected to a simple relation within the Weyl model between the
rate and the current density, as was shown recently \cite{Dora10}.

\section{VII. Discussion, conclusions and generalizations}

To summarize, we studied the dynamics of the electron - hole pair creation
by an electric field in a single graphene sheet where the chemical potential
is located right between the valence and the conduction bands. We assumed
that the transport is purely ballistic, thus neglecting impurities,
interaction with phonons, ripplons, as well as the screened interaction
between electrons. The dynamical approach which originated in particle
physics \cite{Gitman} was adapted to the tight binding model of graphene and
allowed us to calculate the evolution of the current density, energy and
number of pairs beyond linear response. We clarified several delicate issues
within linear response like the correct dc conductivity value and a proper
use of the ultra - relativistic Weyl model approximation to the tight
binding model. The question of proper regularization within the Weyl
approximation was linked to the chiral (parity) symmetry and the anomaly
cancellation in the tight binding model. Using proper regularization the
leading correction to the linear response in both dc and ac fields was
calculated. The ac response which is purely pseudo-dissipative (no inductive
part) and frequency independent in linear response shows strong frequency
dependence, third harmonic generation and the inductive behaviour.

It should be emphasized that beyond linear response new scales appear.
Generally the tight binding model in electric field has a time scale $%
t_{\gamma }=\hslash /\gamma $ \textit{and} a dimensionless parameter $%
\mathcal{E}=E/E_{0}$, $E_{0}=\gamma /\left( ea\right) $. In linear response
the dimensionless parameter $\mathcal{E}$ does not appear in conductivity.
This explains why the ac conductivity is independent of frequency all the
way from optical frequencies $1/t_{\gamma }$ down to dc. Therefore it is not
surprising that the dc conductivity is $\sigma _{2}$ rather than $\sigma
_{1},$ which was obtained in numerous calculations over the years \cite%
{GusyninPRB06,Ando02,Beenaker,Katsnelson06}. Some papers, in which both
values are obtained in different limits \cite{Ziegler06}, even raise the
question, what quantity is actually measured in experiments in "bulk"
graphene like those in ref.\cite{AndreiNN08,BolotinPRL08}? Within the
dynamical approach there is no room for ambiguity since no regularizations,
limits or uncontrollable approximations were used. We considered also finite
size effects for periodic boundary conditions and found that the
conductivity converges to $\sigma _{2}$ for sizes of order $W,L\sim 50a$. We
claim therefore that dc conductivity in graphene is a well defined
physically measurable bulk quantity for a sufficiently large sample, $W,L>>a$%
, and cannot have two different values within the same model. In particular
there is no dependence on the aspect ratio $W/L$ for large samples contrary
to the result of the Landauer approach. Therefore it is important to ask
what could go wrong in other approaches to the same quantity in the same
model.

The various approaches leading to this incorrect value $\sigma _{1}$ can be
broadly divided into two classes. One type of calculations involves as the
first step the introduction of disorder as a way to regularize the problem.
At a later stage the disorder strength is taken to zero \cite%
{FradkinPRB86,Ando02}. The calculation generally involves an uncontrollable
resummation of diagrams. In addition to the Lindhart diagram, other diagrams
(infinite series) involving disorder are summed. For a "regular" system this
leads to the Drude formula with the correct limit (infinite conductivity)
when the limit of vanishing disorder is taken. In graphene the nature of
conductivity is different. There are no charge carriers and the electric
field first has to create the carriers (electron - hole pairs) and only then
to accelerate them. The acceleration is also quite specific: the absolute
value of velocity is fixed, while only the angle can change. The diagrams
omitted in this process might not be small. There is no small parameter like
the Ioffe - Regel $\hbar /\varepsilon _{F}\tau $, where $\tau $ is the
relaxation time, as in the "regular" case. One argues that due to "large" $%
\varepsilon _{F}$ the other diagrams involving crossings are small. At the
Dirac point $\varepsilon _{F}=0$ and this argument should be questioned.
Therefore the use of the simplest resummation might be the origin of the
error.

The second independent approach giving the same value $\sigma _{1}$
originates in mesoscopic physics. Disorder does not appear here and the only
input is the description of "leads" and the boundary effects of the sample.
Within the Landauer approach \cite{Beenaker} one counts the "evanescent"
modes in a ribbon of finite width $W$ and length $L$ leading to a result
(for the armchair boundary conditions) 
\begin{equation}
\sigma =4\frac{e^{2}}{h}\frac{L}{W}\sum_{n=0}\cosh ^{-2}\left( \pi
nW/L\right) \underset{W/L\rightarrow \infty }{\rightarrow }\sigma _{1}\text{.%
}  \label{Beenakker}
\end{equation}%
It was claimed that The predictions including the finite sample conductivity
and the Fano factor were confirmed by experiments \cite{LandauerExp} for
very short samples on substrate. Note that the expression, Eq.(\ref%
{Beenakker}), depends on the aspect ratio only, even when both $W$ and $L$
are large. We have performed calculations for finite samples with periodic
boundary conditions (limited to $1/4<W/L<4$) and the results have a
consistent large size limit of $\sigma _{2}$ irrespective of the aspect
ratio. Thus there is a discrepancy between the two methods which should be
further clarified.

Beyond the linear response the dimensionless parameter $\mathcal{E}$ in
principle can give rise to "macroscopic" time scales $t=\mathcal{E}^{\alpha
}t_{\gamma }$. A rather unexpected result is that at a scale,%
\begin{equation}
t_{nl}\simeq \frac{t_{\gamma }}{\sqrt{\mathcal{E}}}\simeq \sqrt{\frac{\hbar 
}{eEv_{g}}}\text{,}  \label{t_nl}
\end{equation}%
the physical behaviour qualitatively changes. This time scale becomes the
same as the ballistic time $t_{bal}=L/v_{g}=2\cdot 10^{3}t_{\gamma }$ for
length $L=0.5\mu m$ for relatively weak fields $\mathcal{E}=10^{-6}$\
corresponding to $E=10^{4}V/m$. It is important to note that this is the
only time scale that enters the Weyl model approximation to the tight
binding. Within this model the microscopic scale $t_{\gamma }$ does not
appear and $t_{nl}$ is the only combination of the available parameters $%
v_{g}$ and $E$. One could anticipate that the same scale appears in Weyl
model as well \cite{Dora10}. Larger scales however can be formed in the
tight binding model. For example at a scale $t_{B}=t_{\gamma }/\mathcal{E}$
Bloch oscillations set in. This is already beyond the Weyl approximation.

We therefore summarize the behaviour of the conductivity and the number of
electron - hole pairs in the various regimes at finite electric field $E$ in
turn.

\textit{(i)} \textit{Ballistic times} $t<t_{nl}$. \textit{Pseudo -
dissipative behaviour.}

After a brief transient period (of order of several $t_{\gamma }=\hslash
/\gamma $) the current density of the tight binding model approaches a
finite value and stabilizes there. Therefore the minimal dc electric
conductivity at zero temperature is $\sigma _{2}=\frac{\pi }{2}\frac{e^{2}}{h%
}$. The pairs are created with various orientations of velocity (the value
of which is fixed at $v_{g}$) and part of the current (the polarization or
"zitterbewegung" part) is due to reorientation of the charge carriers. In
this regime the pair creation rate given by Eq.(\ref{N_pert1}) is small
enough so that pairs can be considered independent, provided the number
density of created pairs $\frac{e^{2}E^{2}\gamma }{\hslash ^{2}}t^{2}$ is
not too large. If it is very large the inverse process of pair annihilation
(via various channels) cannot be neglected. Approaching $t_{nl}$ the
perturbation theory breaks down. At this time the density of pairs is $\frac{%
eE}{a}$. It becomes of the order of $10^{11}cm^{-1}$ for $E=10^{5}V/m$. Let
us assume that this does not happen and proceed to the longer ballistic
times.

\textit{(i) Ballistic times} $t_{nl}<t<t_{B}$. \textit{Schwinger's pair
creation and formation of the electron - hole plasma.}

On the larger time scale nonperturbative methods should be used. The pair
creation is more intensive and the pair density asymptotically follows an
analog of the Schwinger's formula for massless fermions in 2+1 dimensions,
Eq.(\ref{Schroedinger1}). The current above $t_{nl}$ increases linearly with
time 
\begin{equation}
J\left( t\right) =\sigma _{2}\left( \frac{\sqrt{3}}{2}E\right) ^{3/2}\left( 
\frac{ev_{g}}{\hbar }\right) ^{1/2}t\text{.}  \label{J_nl}
\end{equation}%
In this regime most of the electrons are oriented along the field direction,
so that $J=2ev_{g}N\left( t\right) $. For example, in order to reach the
density of $N=10^{11}cm^{-2}$ in a sample with ballistic time $%
t_{b}=L/v_{g}=2000t_{\gamma }$ ($L=0.5\mu m$) the field should reach $%
E=10^{4}V/m$. If the sample is short enough the transport still can be
ballistic, but due to its nonlinear nature a more likely scenario is that a
dissipation channel opens up.

One can only speculate what kind of dissipation process truncates the pair
creation and stabilizes the electron - hole plasma created by the electric
field. Of course, the standard candidates are collisions with impurities,
phonons, ripplons and the electron - electron interactions. Here we point
out that the system is "open" and one should consider the "radiation
friction" scenario: pairs annihilate emitting photons which take energy out
of the graphene sheet. The effects of radiation of energy into space might
in principle be observable at elevated fields and should be investigated.

\textit{Ballistic times} $t>t_{B}$. \textit{Bloch oscillations.}

If the system is still ballistic at yet longer times, Bloch oscillations set
in with a period of $t_{B}=\frac{\hslash }{eaE}=t_{\gamma }/\mathcal{E}$.
This time scale becomes the same as the ballistic time $t_{bal}=2\cdot
10^{3}t_{\gamma }$ for relatively weak fields $\mathcal{E}=10^{-3}$\
corresponding to $E=10^{7}V/m$, see Fig. 2 in ref. \cite{Rosenstein10}.
While the Bloch oscillations are difficult to observe (see however a recent
proposal \cite{Dragoman08}), the transition to a nonlinear regime is within
reach of current experimental techniques.

The dynamic approach was generalized to bilayer graphene \cite{bilayer} in
which similar questions exist for a long time. The correct dc conductivity
for the $N$ layered graphene is equal to the dynamical one $\sigma =N\sigma
_{2}$, consistent with the vanishing frequency limit of the ac conductivity 
\cite{MacDonald}. The creation of the electron - hole plasma is even more
likely in these systems.

In this paper only ballistic transport was considered. In principle,
disorder and electron - electron interactions could be incorporated within
the dynamical approach in the way the Boltzmann equation approach \cite%
{Sachdev,Auslender} can be extended beyond the linear response. In fact,
using a phenomenological methodology, disorder has recently been
incorporated for the ac field in ref. \cite{MishchenkoPRL09}. Similarly
Coulomb interactions and the pair annihilations into photons, phonons etc.
can be taken into account. Generally though, when the system has a large
number of electron - hole pairs, the screening by the neutral plasma is more
effective and the influence of impurities and interactions diminishes.
Understanding these effects is crucial for investigating the (nonlinear)
plasma waves and their damping \cite{Jena}.

Let us note the relation between the dynamics on the time scale $t_{nl}$ and
quantum adiabatic transport near the quantum critical point \cite%
{Sachdevbook,Santoro}. The calculation of the number of pairs can be
approximately performed using the instanton approach initiated by in the
context of particle physics \cite{Nussinov} and extended in \cite%
{GavrilovPRD96,Cohen}. In condensed matter physics the method is known as
the Landau - Zener probability \cite{Dora10,Santoro}. It provides an
intuitive picture of the Schwinger's pair creation rate. The ballistic
evolution in graphene therefore can be considered as an example of the
adiabatic quantum evolution which attracted much attention recently in
connection to the Kibble - Zurel mechanism of phase transition dynamics and
others. Graphene dynamics at large ballistic times offers an accessible
system in which these processes can be observed.

Finally let us remark on the application of the dynamical approach to
calculating the response to very short strong field pulses like the
femtosecond (and an order of magnitude longer) laser pulses. A possibility
of measuring the response to such fields was advocated by Rusin and Zawadzki 
\cite{Rusin}. For this purpose the dynamical linear response formulas, as
presented in the Section III, can be directly applied for arbitrary time
dependence of the pulse, while the nonlinear calculation of Section V
describes the step - like pulse of finite duration only. Other shapes of the
pulse can be calculated by breaking the pulse into several segments of
different constant field.

Acknowledgements. We are indebted to J. Santoro, E. Andrei, M. Deshmukh, A.
F. Morpurgo, V. Singh, E. Farber and W.B. Jian for valuable discussions.
Work of B.R. and H.K. was supported by NSC of R.O.C. Grants No.
98-2112-M-009-014-MY3 and 98-2112-M-003-002-MY3, respectively, and MOE ATU
program. B.R. acknowledges the hospitality of the Applied Physics Department
of AUCS; M.L. acknowledges the hospitality and support of the NCTS.

\end{document}